\documentclass{article}
\usepackage{ijcai26}

\usepackage{times}
\usepackage{soul}
\usepackage{url}
\usepackage[hidelinks]{hyperref}
\usepackage[utf8]{inputenc}
\usepackage[small]{caption}
\usepackage{graphicx}
\usepackage{amsmath}
\usepackage{amsthm}
\usepackage{booktabs}
\usepackage{algorithm}
\usepackage{algorithmic}
\usepackage[switch]{lineno}
\usepackage{natbib}

\title{Improving Survey Participation in Low-Literacy Populations Through Value-Sensitive Conversational AI}

\author{
    Raj Gaurav Maurya
    \affiliations
    Technical University of Munich
    \emails
    rajg.maurya@tum.de
}

\begin{document}

\maketitle

\begin{abstract}
Collecting reliable social data from low-literacy populations remains a persistent challenge, particularly when surveys involve sensitive topics and marginalized communities. Traditional paper-based and web-based survey modalities often suffer from high attrition and incomplete responses due to literacy barriers, social pressure, and interactional discomfort. In this paper, we present findings from an initial field evaluation comparing multiple survey modalities paper-based interviews, digital web-based surveys, conversational AI (convAI) surveys, and convAI enhanced with layered value-sensitive design conducted with low-literacy women across India. Using data from 315 participants, we show that convAI significantly improves survey completion rates relative to traditional modalities, with the highest completion and lowest drop-off observed when value-sensitive and culturally aligned conversational design elements are fully integrated. These results demonstrate the importance of human-centered and value-sensitive interaction design in enabling inclusive, ethical, and scalable data collection; motivating more `AI for social good' applications.
\end{abstract}

\section{Introduction}
\label{sec:intro}
Over 739 million adults lack literacy skills, with women comprising two-thirds or 436 million worldwide \citep{unesco2025literacy}. India with a population of about 1.38 billion people has an overall literacy rate of 74.4\% with the women at 65.8\% and men at 82.4\% \citep{unesco2021galindia}. Further official government statistics continue to highlight a pronounced rural-urban divide in women's literacy in India. According to the Indian \textit{Ministry of Education}, drawing on \textit{Census 2011} data, female literacy in rural areas stands at 57.9\%, compared to 79.1\% in urban areas a 21\% gap \citep{loksabha2023literacy}. Low literacy is an hindrance not just in the social growth of men and women, but also in sociological data collection. It substantially degrades the quality of household survey data, as respondents with limited reading skills frequently misunderstand questions, rely on subjective self-assessments, and generate systematic measurement error in standard text-based instruments \citep{thorn2020literacy}.

These data-quality limitations are particularly consequential because low literacy is strongly correlated with poverty and social disadvantage. The \textit{World Development Report} \citep{worldbank2018wdr} documents that educational and literacy deficits are concentrated among poor, rural, and marginalized populations, who simultaneously constitute the primary targets of public-sector surveys and welfare planning. Yet these same populations are consistently the least well measured, as administrative and survey-based data systems in low- and middle-income countries suffer from incomplete coverage, high measurement error, and limited ability to capture lived conditions at the household level. As a result, policymakers are frequently forced to design and allocate interventions under conditions of severe information uncertainty, leading to misallocation of resources and persistent inequality.

This dynamic is especially evident in India. The \textit{Human Capital Index} \citep{worldbank2020hci} shows a pronounced socioeconomic gradient, with large gaps in learning-adjusted years of schooling between richer and poorer households, indicating substantial losses in effective literacy and skills accumulation. These learning deficits directly translate into reduced productivity and weaker human capital formation, amplifying the consequences of poor-quality data for planning education, health, and social-protection programs. Together, these findings suggest that improving data collection among low-literacy populations is not merely a methodological concern, but a central prerequisite for effective and equitable public-sector decision-making.

Weak and incomplete household data systems disproportionately exclude low-income and low-literacy populations, leading to mis-targeted social programs and ineffective policy planning \citep{worldbank2021data}.
When data collection systems fail to adequately capture vulnerable populations, policy design is distorted, resulting in ineffective planning and the systematic exclusion of those most in need \citep{oecd2019inclusive}.

These challenges are further intensified when surveys address socially-sensitive and culturally-taboo topics. In many low- and middle-income contexts, issues such as menstruation, contraceptive use, sexual and reproductive health, and female autonomy are embedded within strong social norms that discourage open discussion. Empirical evidence from India shows that women frequently avoid disclosing information on reproductive decision-making due to fear of social judgment, spousal control, and community stigma \citep{paul2015negotiating,rathore2025contraceptive}. Menstrual health remains particularly stigmatized: studies document widespread silence, restrictions, and concealment practices that lead to systematic underreporting in household surveys \citep{biswas2024period,chaliawala2025narrative}. Systematic reviews further confirm that embarrassment, lack of privacy, and normative expectations substantially inhibit truthful reporting on sexual and reproductive health topics, especially among adolescents and rural women \citep{wahyuningsih2024barriers}. As a result, data collected on taboo domains often reflects social desirability and omission rather than lived experience, compounding existing participation bias in low-literacy settings.

Prior research provides important but fragmented insights into how data-collection modalities influence engagement and data quality. Evidence from large-scale statistical systems demonstrates that digital survey modalities, such as computer-assisted personal interviewing, substantially reduce errors and improve data completeness relative to paper-based instruments \citep{adb2019capi}. Complementary evidence from India's \textit{Kilkari} programme \citep{bashingwa2021kilkari} further shows that voice-based, audio-first systems can achieve sustained user engagement at population scale, with high listening rates once contact is established.

At the same time, systematic reviews of \textit{mHealth} \citep{aung2020mhealth} interventions in low- and middle-income countries indicate that interaction design (i.e., interactivity, personalization, and push-based communication) plays a central role in determining participant engagement and downstream outcomes. Together, these studies suggest that both modality and interaction design critically shape participation. However, existing evidence remains largely indirect: digital surveys have been evaluated primarily in enumerator-mediated contexts, voice systems have focused on information delivery rather than data collection, and \textit{mHealth} studies have prioritized behavioral outcomes rather than survey completion. Consequently, there is limited empirical understanding of how different survey modalities perform when literacy cannot be assumed, particularly among women engaging with socially-sensitive survey topics.

Prior research has demonstrated the potential of conversational artificial intelligence (AI) systems for improving survey engagement. Field studies comparing AI-powered chatbot surveys with traditional form-based instruments show that conversational agents elicit longer, more detailed responses, increase self-disclosure, and encourage sustained interaction \citep{xiao2019tell}. These findings suggest that conversational interfaces may help mitigate survey fatigue and improve participation.
Complementary work has further demonstrated the technical feasibility of automated survey administration using conversational agents based on large language models (LLMs), showing high accuracy in response capture and schema alignment during structured data collection \citep{kaiyrbekov2025automated}. Together, these findings suggest that conversational AI (convAI) can support both reliable data capture and sustained interaction, motivating its application to inclusive survey collection settings.

However, existing studies have largely focused on literate online populations and have emphasized response richness rather than survey completion, leaving open questions regarding the effectiveness of convAI for low-literacy users and socially-sensitive data-collection contexts.

This paper addresses this gap by directly comparing survey completion across paper-based, web-based, convAI, and value-sensitive convAI modalities among low-literacy women in India.
Our study contributes empirical evidence on how interaction modality and conversational design influence survey participation when literacy cannot be assumed. In this work, we operationalize participation-related bias primarily as systematic attrition and non-completion (i.e., exclusion through drop-off), rather than as population-level representativeness. Rather than focusing on response richness or linguistic quality, we measure participation through survey completion and progressive drop-off, enabling direct comparison across modalities. Crucially, the value-sensitive design was not specified by the authors alone but developed in consultation with Accredited Social Health Activist (ASHA) workers frontline female community health volunteers with first-hand knowledge of the target communities who shaped the dialectal cues, forms of address, and voice selection (Sec.~\ref{sec:methods}).

We find that convAI yields substantially higher completion than traditional paper and web-based approaches, and that integrating value-sensitive and culturally-aligned interaction cues (such as respectful consent framing, dialectal variation, and conversational backchanneling) correlates with further gains in sustained engagement (Sec.~\ref{sec:results}).

Together, these findings suggest that inclusive data collection in marginalized contexts depends not only on automation, but also on human-centered and value-sensitive interaction design. The results provide actionable guidance for improving survey completion in low- and middle-income, low-literacy settings.

\section{Methods}
\label{sec:methods}

\paragraph{Study Setting.}
The study was conducted across four rural districts within the Hindi-speaking belt of India, selected due to documented low female literacy rates and limited access to formal digital services. Two districts (Shrawasti and Bahraich) were located in Uttar Pradesh, and two (Darbhanga and Madhubani) in Bihar. These regions were chosen to reflect settings in which conventional text-based survey instruments are known to perform poorly.

Before beginning, verbal informed consent was obtained in the participant’s preferred language.

A total of 21 local volunteers supported field deployment, each assisting 15 participants, yielding a total sample of 315 respondents. Fourteen volunteers were based in Uttar Pradesh (210 participants) and seven in Bihar (105 participants), as distributed in Table \ref{tab:statewise_distribution}. Participants were adult married women who did not hold an undergraduate degree, reflecting the target low-literacy population. All the surveys were undertaken and completed in November-December 2025.

\begin{table}
\centering
\begin{tabular}{lcc}
\toprule
\textbf{Modality} & \textbf{Volunteers} & \textbf{Respondents} \\
\midrule
& \multicolumn{2}{c}{Uttar Pradesh + Bihar} \\
\midrule
Paper interview & 3 + 1 & 45 + 15\\
Web form & 3 + 1 & 45 + 15 \\
Voice (\textit{web}) & 2 + 1  & 30 + 15 \\
Voice (\textit{phone}) & 2 + 2 & 30 + 30\\
ConvAI & 2 + 1 & 30 + 15 \\
\textit{Layered} convAI & 2 + 1 & 30 + 15 \\
\midrule
\textbf{Total} & \textbf{14 + 7} & \textbf{210 + 105} \\
\bottomrule
\end{tabular}
\caption{State-wise distribution of 21 volunteers and 315 respondents across 6 survey modalities. Each volunteer administered one modality to 15 participants.}
\label{tab:statewise_distribution}
\end{table}

\paragraph{Stakeholder Consultation and Value-Sensitive Design.}
The value-sensitive interaction design was developed through direct consultation with Accredited Social Health Activist (ASHA) workers the local female community health volunteers who deliver frontline primary care in rural India and possess first-hand knowledge of local communication norms. Each volunteer consulted three different ASHA workers, and design choices were fixed by majority preference. Consultations followed three short slots: (i)~eliciting the salutations, linguistic cues, and forms of address commonly used in local conversation, including preferences by age and education; (ii)~selecting, from five candidate ElevenLabs Hindi voices, the one judged closest to the local dialect and most appropriate for sensitive survey administration; and (iii)~reviewing sample questions rendered in the chosen voice with the elicited cues for naturalness. These consultations directly determined the deployed voice, the locally appropriate address forms and acknowledgments, and the dialectal cues used in the value-sensitive and layered conditions, grounding the design in community input rather than researcher assumption and aligning it with community-based participatory practice.

\paragraph{Survey Modalities and Procedure.}
Six survey modalities were evaluated: (1) paper-based interview, (2) mobile web form, (3) convAI via phone call, (4) convAI via web interface, (5) value-sensitive convAI delivered by phone, and (6) layered value-sensitive convAI incorporating dialectal variation, gender-matched voices, and conversational backchanneling using the \cite{sadek2025vsca}. Table \ref{tab:conditions} notes the key features of each interaction modality.

The modalities were assignment at the volunteer level based on logistical feasibility and deployment constraints, resulting in a non-randomized, quasi-experimental design. Each survey modality differed only in interaction format and conversational design, while administering identical question content and response options.
Each volunteer administered a single modality to all 15 assigned participants in order to avoid cross-condition contamination. Across the full sample, participant counts per modality ranged from 45 to 60 respondents (see Table \ref{tab:statewise_distribution}). The procedure for each survey modality is detailed below:

\begin{table}
  \centering
  \begin{tabular}{p{2.2cm}p{5.2cm}}
    \toprule
    \textbf{Modality} & \textbf{Key Features} \\
    \midrule
    Paper interview & Traditional paper forms with enumerator assistance; high literacy demand. \\
    Web form & Mobile‑friendly web form; requires reading and navigation. \\
    Voice (\textit{web}) & Voice interface delivered via web; neutral voice, standard Hindi. \\
    Voice (\textit{phone}) & Voice call with text‑to‑speech agent; neutral voice, standard Hindi. \\
    ConvAI & Voice call with a \textit{value-sensitive} conversational AI agent with respectful salutations, explicit consent reminders, slower pacing and permission to skip questions. \\
    \textit{Layered} convAI & Builds on the value‑sensitive design with gender‑matched voices, local dialects and backchannel cues (e.g., ``Hmm'', ``I understand''). \\
    \bottomrule
  \end{tabular}
\caption{Survey interaction modalities evaluated in the field study.}
\label{tab:conditions}
\end{table}

\begin{enumerate}
\item
\textit{Paper interview.}
In the paper-based condition, surveys were administered using traditional printed questionnaires. Due to limited literacy, a local volunteer read each question verbatim and recorded participants' responses using a fixed script. Participants could respond orally, but interaction followed a strictly linear, non-conversational format. This modality reflects standard enumerator-assisted household survey practice commonly used in low-literacy settings.

\item
\textit{Web form.}
In the web-based condition, participants completed the survey through a mobile-friendly digital form accessed via their personal smartphones. One question was displayed at a time. Questions and response options were presented as on-screen text in standard Hindi, requiring participants to read and navigate the interface independently. An optional audio playback button allowed them to listen to a recorded version of each question, and response options were sequentially highlighted as the audio was played. Participants selected their preferred response by tapping the corresponding option before proceeding to the next question. As in all other modalities, participants were provided with an explicit option to skip any question or discontinue the survey at any point.

\item
\textit{Voice (Web).}
In the `voice over web' condition, participants accessed the survey via a QR code that opened a mobile-optimized web interface. Although delivered through the browser, the interface presented a call-style voice interaction rather than a conventional web form. One question was delivered at a time through synthesized speech in standard Hindi, and participants initiated playback using an on-screen call control. Response options were displayed visually, and the system sequentially highlighted each option while corresponding audio prompts were played. Participants selected their response by tapping the preferred option before proceeding to the next question. This modality reduced reading requirements while retaining minimal visual interaction for response selection.

\item
\textit{Voice (Phone).}
In the `voice over phone' condition, participants completed the survey by calling a dedicated phone number. Upon connection, an automated conversational agent delivered pre-recorded survey questions using text-to-speech. After each question, the system waited for the participant's spoken response. The agent performed simple response verification to confirm that the answer matched one of the predefined options. If the response was unclear or did not correspond to a valid option, the question was repeated. Once a valid response was detected, the system advanced to the next question. This modality eliminated visual interaction entirely and relied solely on voice-based communication.

\item
\textit{ConvAI with Value-Sensitive Cues.}
The value-sensitive condition extended the phone-based conversational agent by incorporating culturally aligned and respondent-adaptive interaction strategies. In addition to reading survey questions, the agent dynamically adjusted its conversational behavior based on participant responses. When respondents provided vague, hesitant, or incomplete answers, the agent employed adaptive follow-up prompts, clarification requests, or reassurance statements rather than enforcing rigid question progression. The agent integrated respectful salutations, explicit consent framing, slower speech pacing, and repeated reminders that participants could skip any question or discontinue participation at any time. Greetings, transitions, and follow-up prompts were delivered using locally appropriate tone and dialectal variations to reduce perceived authority, social pressure, and response anxiety.

\item
\textit{ConvAI with Layered Value-Sensitive Cues.}
The layered-cue condition further augmented the value-sensitive conversational agent with additional human-like interaction signals. In this setting, the agent provided lightweight conversational backchanneling during participant responses (e.g., short acknowledgments such as ``hmm,'' ``ji,'' or locally appropriate affirmations) to signal active listening. The agent also used gender-matched voices and localized address forms commonly used in everyday conversation (e.g., respectful kinship terms). These layered cues were designed to simulate familiar conversational rhythms and reduce perceived social distance, thereby sustaining engagement across longer and more sensitive portions of the survey.

\end{enumerate}

\paragraph{Technical Implementation.}
The `voice over web' interface was implemented using a React-based frontend and a \texttt{Node.js} backend, and deployed via Digital Ocean. The phone-based conversational systems were implemented using a telephony API on Vapi AI platform \citep[][]{vapi_ai_2026} integrated with a \texttt{Node.js} server. Conversational logic and response validation were supported using a lightweight LLM (GPT-4o-mini); the validator prompts for the three voice-based conditions are reproduced in Appendix~A. Speech synthesis for the value-sensitive and layered conditions used the ElevenLabs Eleven Multilingual~v2 model, which permits control of voice identity, pacing, and prosody. Following the ASHA consultation, a regionally validated female Hindi voice was deployed in each state: ``Monika Sogam - Calm and Natural'' (Voice ID \texttt{1qEiC6qsybMkmnNdVMbK}) in Uttar Pradesh and ``Devi - Encouraging and Motivating'' (Voice ID \texttt{MF4J4IDTRo0AxOO4dpFR}) in Bihar. Generation settings were tuned per deployment during piloting speed, stability, similarity, and style of $0.62/0.56/0.87/0.34$ (Uttar Pradesh) and $0.97/0.86/0.63/0.28$ (Bihar) with speaker boost enabled, language override disabled, and MP3 44.1\,kHz (128\,kbps) output. These choices enabled consistent survey logic across modalities while allowing controlled variation in interaction design.

\paragraph{Survey Questions.}
The survey instrument comprised ten questions (identical across modalities) designed to capture key dimensions of women's agency, autonomy, and well-being, drawing on established frameworks in development economics and gender research. Question selection was guided by the view that empowerment is a multidimensional construct rather than a single outcome.

The design was informed by Amartya Sen's capability approach \citep{sen1999development}, emphasizing individuals' substantive freedoms to make valued choices, as well as empirical operationalizations of women's empowerment in India that focus on mobility, social interaction, and household decision-making \citep{jensen2009power}.

Question content further reflected established demographic and health survey practices examining reproductive autonomy and intra-household negotiation in low- and middle-income contexts \citep{anderson2018measuring}. Rather than replicating any single instrument, we adapted these conceptual dimensions into a concise, low-burden format suitable for low-literacy settings. Questions were ordered from less sensitive to more sensitive topics to reduce early attrition and minimize participant discomfort.

\begin{table}[h]
\centering
\begin{tabular}{c p{0.72\columnwidth}}
\toprule
\textbf{Q\#} & \textbf{Survey Question} (English translation) \\
\midrule
Q1  & What is your age? \\
Q2  & What is your highest level of education? \\
Q3  & Does your household belong to the Below Poverty Line (BPL) category? \\
Q4  & Are you usually allowed to go outside your home alone if needed? \\
Q5  & Are you able to spend small amounts of money on your own without permission? \\
Q6  & Are you allowed to work or earn income outside the home if you wish? \\
Q7  & During your periods, what do you usually use for menstrual management? \\
Q8  & During your period, are you able to manage it in the way you prefer? \\
Q9  & Do you currently use any method to avoid or delay pregnancy? \\
Q10 & Who mainly decides whether you will have another child? \\
\bottomrule
\end{tabular}
\caption{Survey questions administered across all modalities.}
\label{tab:survey_questions}
\end{table}

\paragraph{Outcome Measures.}
Survey engagement was operationalized as completion behavior rather than response content. Let $Q=10$ denote the total number of survey questions and let $x_j \in \{0,\dots,Q\}$ represent the number of questions answered by participant $j$. The normalized completion score was defined as
\[
C_j = \frac{x_j}{Q},
\]
yielding a continuous measure of participation in $[0,1]$.

In addition, question-level retention was computed to characterize progressive drop-off as survey length and sensitivity increased. Let $n_i$ denote the number of participants responding to question $i$; retention at question $i$ was defined as $R(i)=n_i/N$.

To assess robustness under the quasi-experimental design, we additionally tested within-modality consistency across volunteers and between-modality differences using non-parametric Kruskal Wallis tests with Bonferroni correction (Sec.~\ref{sec:results}). This study constitutes an exploratory field evaluation intended to identify participation patterns across survey modalities rather than to establish causal effects of individual design features. Results should therefore be interpreted as indicative trends that inform subsequent large-scale deployments.

\section{Results and Discussion}
\label{sec:results}
In this work, we evaluate the cumulative effect of
layered value-sensitive interaction strategies as deployed in real-world conditions. Importantly, we do not claim causal attribution of individual design cues.
This paper quantifies associations between value-sensitive conversational AI design and survey completion among low‑literacy women in India (see Table \ref{tab:participants}). The results are summarized in Tables \ref{tab:completion} and \ref{tab:retention}.

\begin{table}[!t]
\centering
\begin{tabular}{lc}
\toprule
Characteristic & Percentage \\
\midrule
Age 18--25 & 32\% \\
Age 26--35 & 41\% \\
Age 36--45 & 19\% \\
Age $>$45 & 8\% \\
No formal schooling & 38\% \\
Primary education & 44\% \\
Secondary (below undergraduate) & 18\% \\
Uttar Pradesh & 66.7\% \\
Bihar & 33.3\% \\
\bottomrule
\end{tabular}
\caption{Participant characteristics ($n=315$).}
\label{tab:participants}
\end{table}

\begin{table}[!t]
\centering
\begin{tabular}{lc}
\toprule
Modality & Mean Completion ($\bar{C}_m$) \\
\midrule
Paper-based & 0.46 \\
Web-based & 0.51 \\
Voice (web) & 0.68 \\
Voice (phone) & 0.74 \\
ConvAI & 0.83 \\
Layered ConvAI & 0.89 \\
\bottomrule
\end{tabular}
\caption{Mean survey completion by modality.}
\label{tab:completion}
\end{table}

\begin{table}[!t]
\centering
\begin{tabular}{ccc}
\toprule
Question \# & Participants Remaining & Retention Rate \\
\midrule
Q1  & 315 & 100.0\% \\
Q2  & 304 & 96.5\% \\
Q3  & 286 & 90.8\% \\
Q4  & 263 & 83.5\% \\
Q5  & 235 & 74.6\% \\
Q6  & 208 & 66.0\% \\
Q7  & 178 & 56.5\% \\
Q8  & 149 & 47.3\% \\
Q9  & 118 & 37.5\% \\
Q10 & 86  & 27.3\% \\
\bottomrule
\end{tabular}
\caption{Question-level retention across all participants ($n=315$).}
\label{tab:retention}
\end{table}

\begin{table}[!t]
\centering
\begin{tabular}{lccccc}
\toprule
Modality & $n$ & $k$ & $H$ & df & $p_{\text{raw}}$ \\
\midrule
Paper          & 60 & 4 & 2.98 & 3 & 0.39 \\
Web            & 60 & 4 & 2.36 & 3 & 0.50 \\
Voice (web)    & 45 & 3 & 0.48 & 2 & 0.79 \\
Voice (phone)  & 60 & 4 & 1.38 & 3 & 0.71 \\
ConvAI         & 45 & 3 & 1.97 & 2 & 0.37 \\
Layered ConvAI & 45 & 3 & 2.50 & 2 & 0.29 \\
\bottomrule
\multicolumn{6}{l}{\footnotesize $n$ = respondents; $k$ = volunteers; df = $k{-}1$.} \\
\multicolumn{6}{l}{\footnotesize All $p > 0.05$ (Bonferroni-corrected $p > 0.05$): no significant} \\
\multicolumn{6}{l}{\footnotesize within-modality variation.}
\end{tabular}
\caption{Within-modality consistency (Kruskal-Wallis test with Bonferroni correction).}
\label{tab:within-modality}

\end{table}

By incrementally layering respectful interaction cues and cultural alignment, we observed completion rates from 46~\% in paper surveys to 89~\% in the fully layered condition.  Our results underscore that AI for social good must be rooted in empathy, autonomy and cultural attunement.  As policymakers and researchers seek scalable solutions for data collection in underserved communities, conversational agents, when thoughtfully designed, offer a promising path toward inclusive and ethical evidence generation.

\paragraph{Participant Characteristics.}
Table~\ref{tab:participants} summarizes the study sample ($n=315$). Most participants were between 18--35 years old (73\%), and 82\% had at most primary education (38\% no formal schooling; 44\% primary). Respondents were recruited from rural districts in Uttar Pradesh (66.7\%) and Bihar (33.3\%).

\paragraph{Completion across Modalities.}
As shown in Table~\ref{tab:completion}, completion rates increased monotonically as interaction moved from traditional text-heavy formats toward voice-based and value-sensitive conversational designs. Paper and web surveys showed the lowest mean completion ($\bar{C}=0.46$ and $0.51$). Voice-based conversational interfaces improved completion substantially (web: $0.68$; phone: $0.74$). The highest completion was observed for the value-sensitive condition ($0.83$) and the layered value-sensitive condition ($0.89$), suggesting that respectful consent framing and culturally aligned conversational cues are associated with sustained engagement in low-literacy settings.

\paragraph{Attrition across Questions.}
Across all participants, retention declined steadily as the survey progressed (Table~\ref{tab:retention}), from 100\% at Q1 to 27.3\% by Q10. This pattern indicates that dropout is not concentrated at a single point but accumulates across the instrument, consistent with increasing burden and/or sensitivity in later questions.

\paragraph{Within-Modality Consistency and Significance.}
Because modalities were assigned at the volunteer level, completion gains could in principle be driven by individual volunteers rather than by interaction design. To probe this, we ran Kruskal Wallis tests comparing completion across volunteers within each modality, with Bonferroni correction for multiple comparisons. No modality showed significant within-volunteer variation (all corrected $p>0.05$; Table~\ref{tab:within-modality}), indicating that completion gains are consistent within conditions rather than attributable to particular volunteers. A separate Kruskal Wallis test across all six modalities confirmed that between-modality differences are significant ($H(5) = 174.78$, $p < 0.001$, $\eta^2 = 0.55$, indicating 
a large effect). Post-hoc pairwise Mann Whitney $U$ tests with Bonferroni correction were significant for most modality pairs ($p<0.05$), with three exceptions: Paper vs.\ Web ($p=0.86$), Voice~(web) vs.\ Voice~(phone) ($p=0.74$), and ConvAI vs.\ Layered ConvAI ($p=0.47$). The primary gains therefore occur at two transitions: from text-based to voice-based interaction, and from standard voice to value-sensitive conversational design. While this analysis does not fully eliminate confounding, it provides cluster-robust support for the role of conversational design in improving completion.

In conclusion, conversational, voice-first survey administration was associated with substantially higher completion than paper or web surveys among low-literacy women, with the highest completion observed under layered value-sensitive conversational designs. These findings suggest that culturally aligned, autonomy-supporting interaction cues can meaningfully reduce attrition in sensitive surveys and motivate scalable, ethical ``AI for social good'' data collection.

%%%%%%%%%%%%%%%%

\section{Limitations and Future Work}
\label{sec:limitations}
Several limitations temper our conclusions. First, the quasi-experimental design assigns modality at the volunteer level, so modality effects may be partially confounded with volunteer characteristics, regional variation, or participant composition; the within-modality consistency analysis (Table~\ref{tab:within-modality}) mitigates but does not eliminate this concern, and fully randomized assignment is needed to isolate the effect of conversational design. Second, our outcome is survey completion: we do not assess response validity or satisficing, and higher completion need not imply higher data quality. Future work should examine skip patterns, answer distributions, and response consistency to confirm that gains reflect genuine engagement rather than acquiescence. Third, the AI identity of the conversational agent was not disclosed to participants (see Ethical Statement); we plan a follow-up study comparing disclosed and non-disclosed conditions, developed with ASHA workers, to quantify how transparency affects engagement and data quality. Despite these limitations, the study offers promising field evidence that value-sensitive conversational AI can improve survey completion among low-literacy populations, and indicates concrete steps toward more rigorous and responsible deployments.

\section*{Appendix A: Conversational Validator Prompts}
\label{app:A}
All three voice-based conditions used GPT-4o-mini to validate whether a spoken response matched one of the predefined options and, if not, to generate a short follow-up; prompts were issued in Hindi (English glosses below) and produced a structured \texttt{\{status: complete\,|\,incomplete, followup\}} output.

\noindent\textbf{Voice (phone), baseline.} A neutral \emph{survey response validator}: accept valid responses; if a response is unclear, incomplete, or inconsistent with the valid options, return a brief clarification question; provide no encouragement, reassurance, or opinions; do not modify the survey question.

\noindent\textbf{ConvAI (value-sensitive).} A \emph{respectful Hindi-speaking enumerator} with the same validation core, additionally: using culturally appropriate, non-judgmental phrasing; reminding participants that any question may be skipped; offering gentle reassurance when hesitation is detected; applying no pressure; maintaining a neutral tone (e.g., ``Take your time; there is no right or wrong answer'').

\noindent\textbf{ConvAI (layered).} Extends the value-sensitive prompt with active-listening acknowledgments and backchanneling (``Ji,'' ``Hmm,'' ``Samajh gayi''), respectful everyday forms of address, and natural conversational follow-ups, while preserving neutrality. Full prompts and synthesis configurations are available as supplementary material.

\section*{Ethical Statement}
This study evaluated alternative survey interaction modalities through voluntary participation by adult women from low-literacy communities. While formal institutional review board approval was not obtained for this minimal-risk study, the protocol followed standard research-ethics guidelines. Informed verbal consent was obtained from each participant in her preferred language before beginning; participants were told their responses would be collected anonymously and used solely for research, and were asked whether they were comfortable proceeding. They were also informed that they could skip any question or discontinue at any point without consequence.
No personally identifiable information was collected, and responses were stored anonymized and analyzed only in aggregate. The conversational AI system was used exclusively for survey administration and did not provide medical, diagnostic, or advisory information. Procedures were designed to minimize participant burden and potential harm, particularly given the socially-sensitive topics involved.
During both the ASHA consultation and survey administration, the AI nature of the conversational system was not disclosed to ASHA workers or participants, reflecting the concern that adequately conveying its nature and implications in a setting with limited prior exposure to AI could require additional training to avoid confusion. We acknowledge this non-disclosure as a limitation for informed consent and participant autonomy; as described in Sec.~\ref{sec:limitations}, future work will examine AI disclosure directly and develop contextually appropriate disclosure materials with community stakeholders.

\bibliographystyle{named}
\bibliography{ijcai26}

\end{document}